\documentclass[10pt]{article}
\usepackage{graphicx}
\usepackage{amsmath}
\usepackage{amssymb}
\usepackage{caption2}
\begin{document}
\bibliographystyle{prsty}
\begin{center}
{\large {\bf \sc{  Comment on "$Z_c$-like spectra from QCD Laplace sum rules at NLO" }}} \\[2mm]
Zhi-Gang  Wang \footnote{E-mail: zgwang@aliyun.com.  }   \\
 Department of Physics, North China Electric Power University, Baoding 071003, P. R. China
\end{center}

\begin{abstract}
In Phys. Rev. {\bf D103} (2021) 074015 (arXiv:2101.07281), huge (or largest) mass gaps  between the ground states and first radial excited states
for the tetraquark molecular states are obtained, $\Delta=1.70\sim 2.35\,\rm{GeV}$.
We explore the problems in the operator product expansion, Borel windows,  etc, in details.
\end{abstract}

In the QCD, we deal with the non-perturbative vacuum, and have to resort to non-zero vacuum expectation values of the normal-ordered
  quark-gluon operators to describe the hadron properties in a satisfactory way.
   We usually parameterize the vacuum matrix elements  in terms of
 \begin{eqnarray}
 \langle0|:\bar{q}_{\alpha}^{i} q_{\beta}^{j}:|0\rangle &=&\frac{1}{4N_c}\langle0|:\bar{q} q:|0\rangle \delta_{ij}\delta_{\alpha\beta}\, ,
 \end{eqnarray}
 \begin{eqnarray}
 \langle0|:\bar{q}_{\alpha}^{i}  q_{\beta}^{j}G^a_{\mu\nu}:|0\rangle &=&
 \frac{1}{24(N_c^2-1)}\langle0|:\bar{q} g_s\sigma\cdot G q:|0\rangle\left(\sigma_{\mu\nu}\right)_{\beta\alpha}\frac{\lambda^a_{ji}}{2}\, ,
 \end{eqnarray}
  \begin{eqnarray}
 \langle0|:\bar{q}_{\alpha}^{i} q_{\beta}^{j}\bar{q}_{\lambda}^{m} q_{\tau}^{n}:|0\rangle&=&\frac{\kappa}{16N_c^2}\langle0|:\bar{q} q:|0\rangle^2
  \left(\delta_{ij}\delta_{mn}\delta_{\alpha\beta}\delta_{\lambda\tau}-\delta_{in}\delta_{jm}\delta_{\alpha\tau}\delta_{\beta\lambda} \right)\, ,\nonumber\\
  &=&\frac{1}{16N_c^2}\langle0|:\bar{q} q\bar{q} q:|0\rangle
 \left(\delta_{ij}\delta_{mn}\delta_{\alpha\beta}\delta_{\lambda\tau}-\delta_{in}\delta_{jm}\delta_{\alpha\tau}\delta_{\beta\lambda} \right)\, ,
  \end{eqnarray}
 etc, where the $i$, $j$, $m$ and $n$ are color indexes, the $\alpha$, $\beta$, $\lambda$ and $\tau$ are Dirac spinor indexes, the $\lambda^a$ are the Gell-mann matrixes.
 Except for the
 quark condensates, which indicate  spontaneous breaking of the Chiral symmetry through the Gell-Mann-Oakes-Renner relation
 $f^2_{\pi}m^2_{\pi}=-2(m_u+m_d)\langle0|:\bar{q} q:|0\rangle$ \cite{GMOR}, other vacuum condensates, such as $\langle0|:\bar{q} g_s\sigma\cdot G q:|0\rangle$,
 $\langle0|:\bar{q} q\bar{q} q:|0\rangle$, $\kappa\langle0|:\bar{q} q:|0\rangle^2$, $\cdots$ are just parameters introduced by hand to describe the non-perturbative vacuum,
 we can parameterize the non-perturbative properties  in one way or the other.

According to the arguments of Shifman,  Vainshtein and  Zakharov and Ioffe \cite{SVZ79,Ioffe-SB}, the accuracy of factorization hypothesis is of order
$\frac{1}{N_c^2}$, the vacuum saturation (factorization) works well in the large $N_c$ limit $\frac{1}{N_c^2}\sim 0$ \cite{Novikov--shifman}, in reality,
$N_c = 3$, $\frac{1}{N_c^2}\sim 10\%$, {\bf it is obvious that the value $\kappa=3\sim4$ chosen in Phys. Rev. {\bf D103} (2021) 074015 is too large}.

In the QCD sum rules for the heavy-light mesons, we often choose the currents $ J_{\Gamma}(x)$ or $J_{\Gamma^\prime}(x)$,
\begin{eqnarray}
 J_{\Gamma}(x)&=& \bar{Q}(x)\Gamma q(x)\, , \nonumber\\
J_{\Gamma^\prime}(x)&=& \bar{q}(x)\Gamma^{\prime}Q(x)\, .
\end{eqnarray}
to interpolate the heavy-light mesons, where the $\Gamma$ and $\Gamma^\prime$ are some Dirac matrixes, and resort to the two-point correlation functions
$\Pi_{\Gamma/\Gamma^\prime}(p)$,
\begin{eqnarray}
\Pi_{\Gamma/\Gamma^\prime}(p)&=&i\int d^4x e^{ip \cdot x} \langle0|T\Big\{J_{\Gamma/\Gamma^\prime}(x)J^{\dagger}_{\Gamma/\Gamma^\prime}(0)\Big\}|0\rangle \, ,
\end{eqnarray}
to study the masses and decay constants.

We usually carry out the operator product expansion up to the vacuum condensates of dimension $6$,
\begin{eqnarray} \label{OPE-D-meson}
\Pi_{\Gamma/\Gamma^\prime}(p)&=&C_0^{\Gamma/\Gamma^\prime}+C_{\bar{q}q}^{\Gamma/\Gamma^\prime} \langle \bar{q}q\rangle
+C_{GG}^{\Gamma/\Gamma^\prime} \langle \frac{\alpha_s}{\pi}GG\rangle+C_{\bar{q}Gq}^{\Gamma/\Gamma^\prime} \langle \bar{q}g_s\sigma Gq\rangle
+C_{\bar{q}q^2}^{\Gamma/\Gamma^\prime}\kappa \alpha_s \langle \bar{q}q\rangle^2\, ,
\end{eqnarray}
where the $C_0^{\Gamma/\Gamma^\prime}$, $C_{\bar{q}q}^{\Gamma/\Gamma^\prime}$, $C_{GG}^{\Gamma/\Gamma^\prime}$, $C_{\bar{q}Gq}^{\Gamma/\Gamma^\prime} $
and $C_{\bar{q}q^2}^{\Gamma/\Gamma^\prime}$ are Wilson's coefficients, the parameter $\kappa$ parameterizes  deviation from  the vacuum saturation hypothesis
 \cite{Narison-PLB-2005,Khodjamirian}, while in Ref.\cite{WZG-Decay-constant}, we assume vacuum saturation and
also take account of the three-gluon condensate $\langle g_s^3 GGG\rangle$.
The accuracy of factorization hypothesis is of order
$\frac{1}{N_c^2}\sim 10\%$, where $N_c = 3$ is the number of colors \cite{Ioffe-SB}. In the QCD sum rules for the traditional mesons,
 the $\langle\bar{q}q\rangle^2$ is always companied with the fine-structure constant $\alpha_s$, see Eq.\eqref{OPE-D-meson} for example, and plays a minor important role,
the deviation from $\kappa=1$ cannot make much difference, although the value $\kappa>1$ can lead to better QCD sum rules in some cases \cite{Review-rho-kappa}.

In the QCD sum rules for the hidden-charm or hidden-bottom tetraquark molecular states, we often choose the color-singlet-color-singlet type local four-quark currents,
\begin{eqnarray}
 J(x)&=& J_{\Gamma}(x)J_{\Gamma^\prime}(x)\, ,
\end{eqnarray}
to interpolate the tetraquark molecular states, and resort to the two-point correlation functions $\Pi(p)$,
\begin{eqnarray}
\Pi(p)&=&i\int d^4x e^{ip \cdot x} \langle0|T\Big\{J(x)J^{\dagger}(0)\Big\}|0\rangle \, ,
\end{eqnarray}
to study the masses and decay constants (or pole residues).

At the QCD side of the correlation functions, we can make a rough estimation about  how to truncate the vacuum condensates in the operator product expansion,
\begin{eqnarray}
\Pi(p)&\propto& \Pi_{\Gamma}(p) \otimes\Pi_{\Gamma^\prime}(p)+\cdots \, , \nonumber\\
&\propto& \cdots +C_{\bar{q}q^2}\langle \bar{q}q\rangle^2+ C_{\bar{q}q\bar{q}Gq} \langle \bar{q}q\rangle\langle \bar{q}g_s\sigma Gq\rangle
+ C_{\bar{q}Gq^2}  \langle \bar{q}g_s\sigma Gq\rangle^2+\cdots\, ,
\end{eqnarray}
where the $C_{\bar{q}q^2}$, $C_{\bar{q}q\bar{q}Gq}$ and $C_{\bar{q}Gq^2}$ are the Wilson's coefficients. In the QCD sum rules for the heavy-light mesons, the quark condensate and mixed condensate play
an important role \cite{Narison-PLB-2005,Khodjamirian,WZG-Decay-constant}, so in the QCD sum rules for the tetraquark molecular states,
we should take account of the vacuum condensates $\langle \bar{q}q\rangle^2$,
$\langle \bar{q}q\rangle\langle \bar{q}g_s\sigma Gq\rangle$ and $\langle \bar{q}g_s\sigma Gq\rangle^2$ at least.

\begin{figure}
 \centering
  \includegraphics[totalheight=5cm,width=7cm]{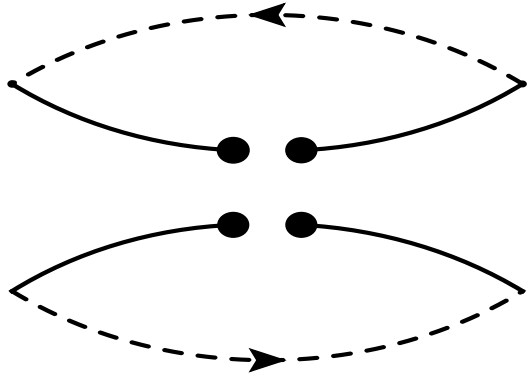}
    \caption{     The  Feynman diagram  contributes  to  the    $\kappa\langle\bar{q}q\rangle^2$,
     where the solid lines and dashed lines denote the light quarks and heavy quarks, respectively.   }\label{OPE-qq-qq}
\end{figure}

\begin{figure}
 \centering
 \includegraphics[totalheight=5cm,width=14cm]{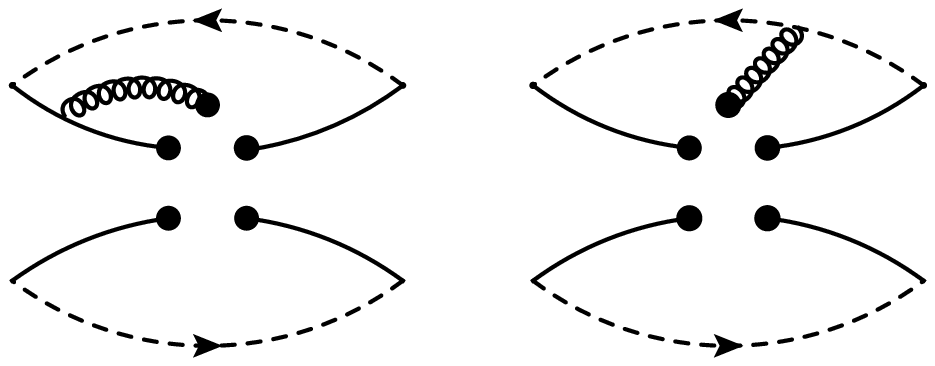}\\
 \vspace{1cm}
  \includegraphics[totalheight=5cm,width=14cm]{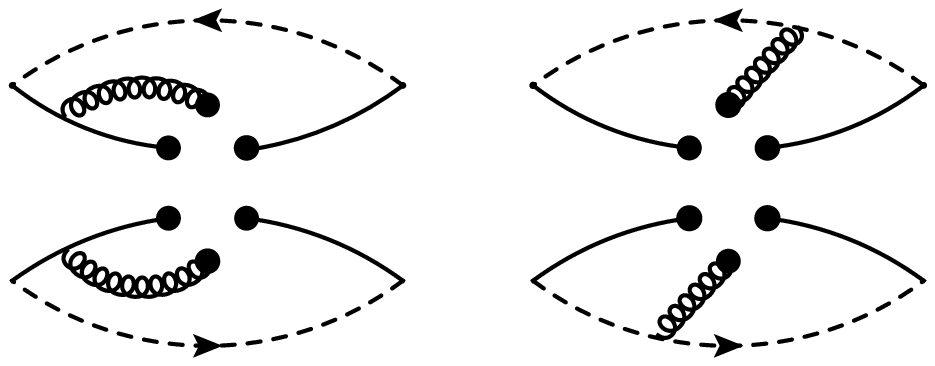}
    \caption{ The  Feynman diagrams  contribute  to  the    $\langle\bar{q}q\rangle\langle\bar{q}g_s\sigma Gq\rangle$
    and $\langle\bar{q}g_s\sigma Gq\rangle^2$. Other diagrams obtained by interchanging of the light  and  heavy quark lines are implied.
       }\label{OPE-qqg-qqg}
\end{figure}

In Phys. Rev. {\bf D103} (2021) 074015, the operator product expansion is accomplished up to the vacuum condensates of dimension 6 \cite{Narison-2101.07281}.
The contributions of the vacuum condensates $\kappa\langle\bar{q}q\rangle^2$ can be shown diagrammatically  in Fig.\ref{OPE-qq-qq}.
If the deviation  from  the vacuum saturation is large, for example, $\kappa= 3\sim4$ in Phys. Rev. {\bf D103} (2021) 074015 \cite{Narison-2101.07281}, which means
that the contributions of the four-quark condensates are amplified about $3\sim 4$ times, it is no reason to discard the contributions of the vacuum condensates
$\langle\bar{q}q\rangle\langle\bar{q}g_s\sigma Gq\rangle$ and $\langle\bar{q}g_s\sigma Gq\rangle^2$, which are shown  diagrammatically  in Fig.\ref{OPE-qqg-qqg}.
The Feynman diagrams shown in Fig.\ref{OPE-qq-qq} and Fig.\ref{OPE-qqg-qqg} are analogous, they should be taken into account or discarded simultaneously. The Feynman diagrams shown in Fig.\ref{OPE-qqg-qqg} cannot be discarded
just because "our poor knowledge of their size", "the validity of the vacuum saturation used for its estimate is questionable \cite{Narison-2101.07281}".
In fact, no direct theoretical calculations can improve that: the vacuum condensates $\langle\bar{q}q\rangle^2$ are  associated with a factor $\kappa=3\sim4$, and 
the vacuum condensates $\langle\bar{q}q\rangle\langle\bar{q}g_s\sigma Gq\rangle$ and $\langle\bar{q}g_s\sigma Gq\rangle^2$ are not associated with the same factor
 $\kappa=3\sim4$. {\bf The truncation of the operator product expansion in Ref.\cite{Narison-2101.07281} is too crude to make reliable predictions}.
In Ref.\cite{Lee-EPJA}, H. J. Lee observes that the vacuum condensates $\langle\bar{q}q\rangle\langle\bar{q}g_s\sigma Gq\rangle$ play a curial role in the QCD sum rules
for the light tetraquark states. In the QCD sum rules for the hidden-charm or hidden-bottom tetraquark (molecular) states, we should carry out the
operator product expansion up to the vacuum condensates of dimension $10$ in a consistent way \cite{WZG-CPC-Z4600,WZG-PRD-hideen-charm}, and take account of
the vacuum condensates $\langle\bar{q}q\rangle$, $\langle\frac{\alpha_{s}GG}{\pi}\rangle$, $\langle\bar{q}g_{s}\sigma Gq\rangle$, $\langle\bar{q}q\rangle^2$,
$\langle\bar{q}q\rangle \langle\frac{\alpha_{s}GG}{\pi}\rangle$,  $\langle\bar{q}q\rangle  \langle\bar{q}g_{s}\sigma Gq\rangle$,
$\langle\bar{q}g_{s}\sigma Gq\rangle^2$ and $\langle\bar{q}q\rangle^2 \langle\frac{\alpha_{s}GG}{\pi}\rangle$.

In Phys. Rev. {\bf D103} (2021) 074015, whether or not the operator product expansion is convergent is not checked,  in such an extreme condition that the  contributions
of the four-quark condensates, the  vacuum condensates of the highest dimension,  are amplified about $3\sim 4$ times. In addition, the continuum threshold parameters and
energy scales of the QCD spectral densities are postponed to very large values to obtain energy scale independent QCD sum rules. However,
from {\bf Fig.11-b, Fig.13-b, Fig.14-b, Fig.16-b, Fig.19-b in Ref.\cite{Narison-2101.07281}, we cannot see any Borel platforms to extract the molecule masses}.

In Table \ref{tc-M}, we present the continuum threshold parameters, energy scales of the QCD spectral densities and (ground state) tetraquark molecule masses
obtained in Ref.\cite{Narison-2101.07281}. From the Table, we can see clearly that  $\sqrt{t_c}-M\gg 0.5\sim 0.6\,\rm{GeV}$,
{\bf such large continuum threshold parameters are beyond permission of the QCD sum rules}.

In Table \ref{EnergyGaps}, we present the energy gaps between the ground states and first radial excited states of the conventional mesons from the Particle Data Group
and LHCb collaboration \cite{PDG,LHCb-Ds2S}. From the Table, we can see that the mass gaps are less than $0.7\,\rm{GeV}$ if there exist one or two heavy quarks, while the
largest mass gaps are about $1\,\rm{GeV}$ for the $\pi$ and $K$ mesons with the $J^{PC}=0^{-+}$ due to
the fact that the light pseudoscalar mesons play a double role, as both Goldstone bosons and $q\bar{q}$ bound states \cite{WZG-CPL}.
In Phys. Rev. {\bf D103} (2021) 074015, huge mass gaps between the ground states and first radial excited states of the hidden-charm molecular states are obtained,
\begin{eqnarray}
\Delta &=&1.70\sim2.35\,\rm{GeV}\, ,
\end{eqnarray}
{\bf  the largest mass gaps known up to now, and they are beyond the accommodation of the quark model}.

The $Z_c(3900)$ and $Z_c(4430)$ are usually assigned to be the ground state and first radial excited state of the tetraquark (molecular) states respectively according to the
 analogous decays,
\begin{eqnarray}
Z_c(3900)&\to&J/\psi\pi\, , \nonumber \\
Z_c(4430)&\to&\psi^\prime\pi\, ,
\end{eqnarray}
analogous  mass gaps $M_{Z_c(4430)}-M_{Z_c(3900)}=m_{\psi^\prime}-m_{J/\psi}$ \cite{Z4430-1405,Nielsen-1401,Wang4430}. While $m_{\psi^\prime}-m_{J/\psi}\ll \Delta$.

\begin{table}
\begin{center}
\begin{tabular}{|c|c|c|c|c|c|c|c|}\hline\hline
                          & $\sqrt{t_c}$(GeV)   & $\mu$(GeV)       & $M$(GeV)  &$\sqrt{t_c}-M$(GeV)       \\ \hline

$D^*\bar{D}$              &$4.69-6.16$          &$4.65$            & $3.912$   &$0.78-2.25$        \\ \hline

$D_s^*\bar{D}$            &$4.69-6.16$          &$4.65$            & $3.986$   &$0.70-2.18$       \\ \hline

$D^*\bar{D}_s$            &$4.69-6.16$          &$4.65$            & $3.979$   &$0.71-2.19$        \\ \hline

$D_s^*\bar{D}_s$          &$4.90-6.32$          &$4.65$            & $4.091$   &$0.81-2.23$      \\ \hline

$D_0^*\bar{D}_1$          &$5.29-6.32$          &$4.65$            & $4.023$   &$1.27-2.30$      \\ \hline

$D_{s0}^*\bar{D}_1$       &$5.29-6.63$          &$4.65$            & $4.064$   &$1.23-2.57$      \\ \hline

$D_0^*\bar{D}_{s1}$       &$5.29-6.63$          &$4.65$            & $4.070$   &$1.22-2.56$      \\ \hline

$D_{s0}^*\bar{D}_{s1}$    &$5.29-6.63$          &$4.65$            & $4.198$   &$1.09-2.44$      \\ \hline

\end{tabular}
\end{center}
\caption{ The continuum threshold parameters, energy scales of the QCD spectral densities and ground state molecule masses
taken from Ref.\cite{Narison-2101.07281}.  }\label{tc-M}
\end{table}

\begin{table}
\begin{center}
\begin{tabular}{|c|c|c|c|c|c|c|c|}\hline\hline
       ($J^{PC}$)     & $n=1$(MeV)         & $n=2$ (MeV)       & Energy Gap (MeV)       \\ \hline

$\pi(0^{-+})$         &$139.57$             &$1300$            & $1160.43$         \\ \hline

$K(0^{-+})$           &$493.677$            &$1482.40$         & $988.723$         \\ \hline

$\eta(0^{-+})$        &$547.862$            &$1294$            & $746.138$         \\ \hline

$\eta^\prime(0^{-+})$ &$957.78$             &$1475$            & $517.22$         \\ \hline \hline

$\rho(1^{--})$        &$775.26$             &$1465$            & $689.74$         \\ \hline

$\omega(1^{--})$      &$782.65$             &$1410$            & $627.35$         \\ \hline

$\phi(1^{--})$        &$1019.461$           &$1680$            & $660.539$         \\ \hline\hline

$a_1(1^{++})$         &$1230$               &$1655$            & $425$         \\ \hline

$h_1(1^{+-})$         &$1166$               &$1594$            & $428$         \\ \hline

$a_2(2^{++})$         &$1316.9$             &$1705$            & $388.1$         \\ \hline

$f_2(2^{++})$         &$1275.5$             &$1639$            & $363.5$         \\ \hline \hline

$\rho_3(3^{--})$      &$1688.8$             &$1982$            & $293.2$         \\ \hline \hline

$D_s(0^{-+})$         &$1969.0$             &$2591$            & $622$         \\ \hline\hline

$\eta_c(0^{-+})$      &$2983.9$             &$3637.5$          & $653.6$         \\ \hline

$J/\psi(1^{--})$      &$3096.900$           &$3686.097$        & $589.197$         \\ \hline\hline

$\eta_b(0^{-+})$      &$9398.7$             &$9999.0$          & $600.3$         \\ \hline

$\Upsilon(1^{--})$    &$9460.30$            &$10023.26$        & $562.96$         \\ \hline

$\chi_{b0}(0^{++})$   &$9859.44$            &$10232.5$         & $373.06$         \\ \hline

$\chi_{b1}(1^{++})$   &$9892.78$            &$10255.46$        & $362.68$         \\ \hline

$h_{b}(1^{+-})$       &$9899.3$             &$10259.8$         & $360.5$         \\ \hline

$\chi_{b2}(2^{++})$   &$9912.21$            &$10268.65$        & $356.44$         \\ \hline\hline

$B_c(0^{-+})$         &$6274.9$             &$6871.6$          & $596.7$         \\ \hline\hline

\end{tabular}
\end{center}
\caption{ The energy gaps between the ground states and first radial excited states from the Particle Data Group except for
the mass of the $D_s^\prime$ is taken from the LHCb collaboration \cite{LHCb-Ds2S}. }\label{EnergyGaps}
\end{table}

\section*{Acknowledgements}
This  work is supported by National Natural Science Foundation, Grant Number  11775079.

\end{document}